\documentclass[10pt]{article}

\usepackage[english]{babel}

\usepackage[letterpaper,top=2cm,bottom=2cm,left=3cm,right=3cm,marginparwidth=1.75cm]{geometry}

\usepackage{amsmath}
\usepackage{graphicx}
\usepackage[hidelinks]{hyperref}
\usepackage{titling}
\usepackage{lineno}
\usepackage{subfiles}
\usepackage{fixltx2e}
\usepackage{authblk}

\title{Integrated resonant electro-optic comb enabled by platform-agnostic laser integration}

\author[$1,2,\dag$]{Isaac Luntadila Lufungula}
\author[$3,\dag$]{Amirhassan Shams-Ansari}
\author[$3$]{Dylan Renaud}
\author[$1,2$]{Camiel Op de Beeck}
\author[$1,2$]{Stijn Cuyvers}
\author[$1,2$]{Stijn Poelman}
\author[$1,2$]{Maximilien Billet}
\author[$1,2$]{Gunther Roelkens}
\author[$3,*$]{Marko Loncar}
\author[$1,2,*$]{Bart Kuyken}

\affil[$1$]{Photonics Research Group, Department of information technology (INTEC), Ghent University-imec, 9052 Ghent, Belgium}
\affil[$2$]{Center for Nano- and Biophotonics (NB-Photonics), Ghent University-imec, 9052 Ghent, Belgium}
\affil[$3$]{John A. Paulson School of Engineering and Applied Sciences, Harvard University, Cambridge,MA 02138, Massachusetts, USA}
\affil[$*$]{\small{Corresponding authors: bart.kuyken@ugent.be 	+32-9-264-3335, loncar@seas.harvard.edu (617) 495-5798 }}
\affil[$*$]{\small{Contributing authors: Isaac.LuntadilaLufungula@ugent.be, ashamsansari@g.harvard.edu, renaud@g.harvard.edu, cob@ligentec.com, Stijn.Cuyvers@UGent.be,  Stijn.Poelman@UGent.be, Maximilien.Billet@UGent.be, Gunther.Roelkens@ugent.be }}
\affil[$\dag$]{\small{Authors contributed equally to this work.}}

\date{}

\begin{document}
\maketitle

\begin{abstract} 
The field of integrated photonics has significantly impacted numerous fields including communication, sensing, and quantum physics owing to the efficiency, speed, and compactness of its devices. However, the reliance on off-chip bulk lasers compromises the compact nature of these systems. While silicon photonics and III-V platforms have established integrated laser technologies, emerging demands for ultra-low optical loss, wider bandgaps, and optical nonlinearities necessitate other platforms.
Developing integrated lasers on less mature platforms is arduous and costly due to limited throughput or unconventional process requirements. In response, we propose a novel platform-agnostic laser integration technique utilizing a singular design and process flow, applicable without modification to a diverse range of platforms. Leveraging a two-step micro-transfer printing method, we achieve nearly identical laser performance across platforms with refractive indices between 1.7 and 2.5. Experimental validation demonstrates strikingly similar laser characteristics between devices processed on lithium niobate and silicon nitride platforms. Furthermore, we showcase the integration of a laser with a resonant electro-optic comb generator on the thin-film lithium niobate platform, producing over 80 comb lines spanning 12 nm. This versatile technique transcends platform-specific limitations, facilitating applications like microwave photonics, handheld spectrometers, and cost-effective Lidar systems, across multiple platforms.

\end{abstract}

\section{Introduction}

Integrated photonics has sparked a revolution in various fields, including optical communication \cite{zhang2021integrated,atabaki2018integrating}, quantum systems \cite{pelucchi2022potential,moody20222022}, ranging \cite{doylend2020overview}, sensing \cite{chen2014heterogeneously}, metrology \cite{kippenberg2018dissipative}, radio frequency photonics \cite{marpaung2019integrated}, and intelligent computing \cite{Perez:20}. This transformative wave is driven by the capability of integrated photonics to enable high-speed, cost-effective, power-efficient, and compact devices. 
As a result of this expanding utility and growing demand, there is a conspicuous interest in the integration of lasers into various photonics platforms. This integration facilitates increased complexity, reduces the physical footprint, and ultimately lowers the overall cost of these integrated photonic devices.

Although silicon photonics or III-V platforms like InP have well-developed integrated lasers \cite{Siew:21,duan2001indium}, typically used in telecommunication, other platforms are better suited for applications such as nonlinear photonics, ranging, and spectroscopy. In particular ultra-low optical loss, larger transparency window, larger bandgap, and optical nonlinearities are hard to achieve in the established platforms. If one or more of these characteristics are required, new, often comparatively low index, platforms such as Thin-film Lithium Niobate (TFLN) \cite{zhu2021integrated}, Thin-film Lithium Tantalate (TFLT) \cite{wang2023lithium}, Si\textsubscript{3}N\textsubscript{4} \cite{blumenthal2018silicon}, and tantalum pentoxide \cite{splitthoff2020tantalum} (tantala, Ta\textsubscript{2}O\textsubscript{5}) can offer a solution. Each of these platforms could cater to a wide range of applications with integrated lasers but the design of scalable laser-integrated photonic devices on these platforms presents considerable challenges. 

We note that both hybrid (chip-to-chip) and heterogeneous (on-chip) integration techniques \cite{kaur2021hybrid}, are costly and challenging. This is mainly due to either low throughput or non-conventional process development which is imposed by the specific limitations of each platform. 
For hybrid integration, the fabrication process is simplified by eliminating any co-processing of the laser material and the platform material. Butt coupling is used to couple the light of the laser on one chip to the platform circuit on another chip, enabling pretesting and postselection of laser dies. However, this die-to-die integration results in low throughput and requires a coupling region tailored to each platform to ensure low-loss coupling and stable laser operation. 
Conversely, conventional heterogeneous integration techniques, such as wafer bonding or epitaxial growth, offer high throughput by processing the full laser stack in-situ on the platform chip. This, however, requires developing a new process flow compatible with both the gain medium and the platform resulting in a long development time, and a more complex process flow. In other words, conventional hybrid and heterogeneous integration techniques necessitate defining either the cavity or the coupling interface on the platform material which incurs significant costs due to the need for platform-specific fabrication processes and designs.

\begin{figure}[ht!]
\centering\includegraphics[width=\textwidth]{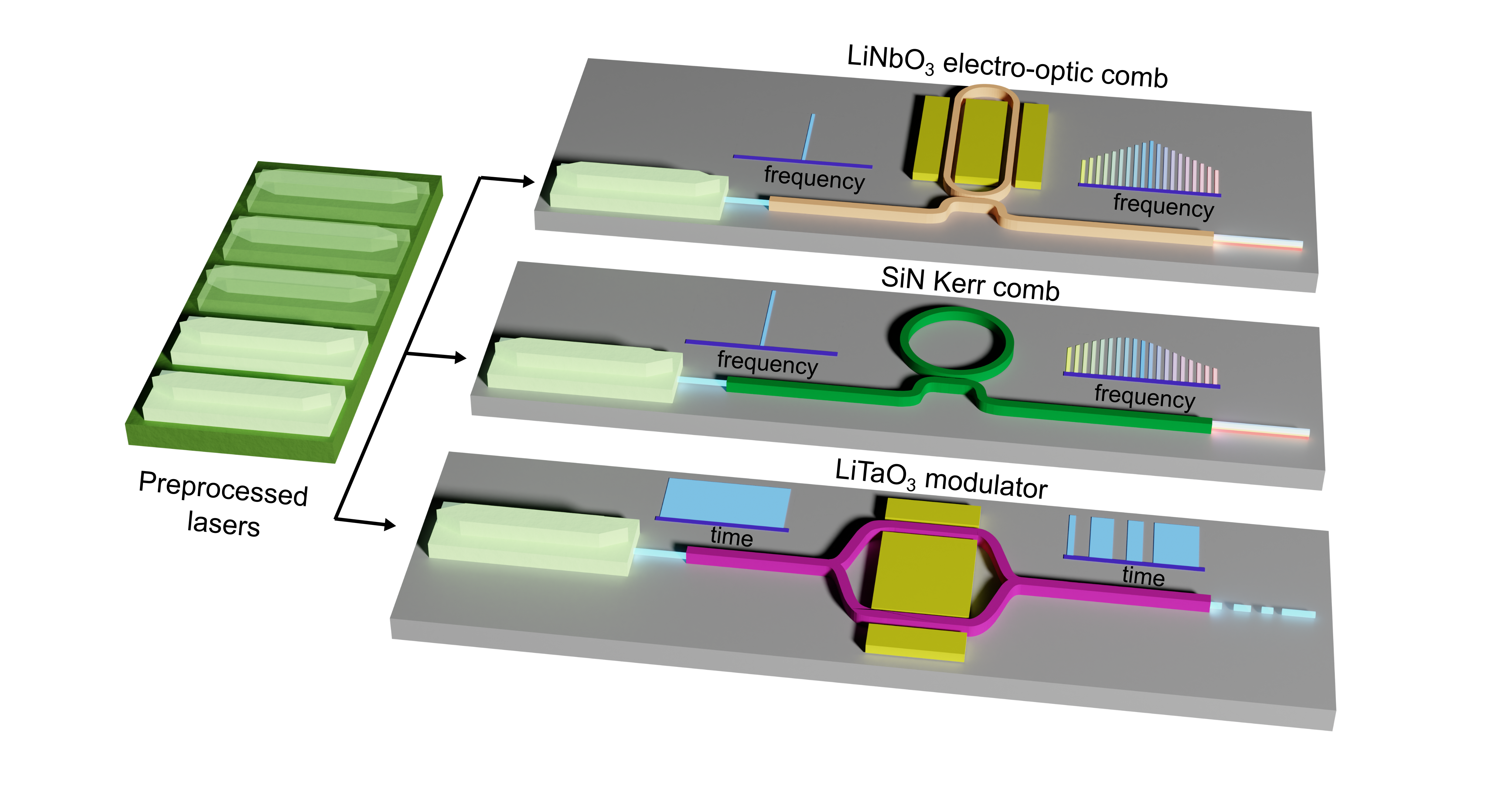}
\caption{\textbf{Conceptual representation of platform-agnostic laser integration:} \\The envisioned platform-agnostic laser integration technique utilizes the same design and process flow for a wide range of nanophotonic platforms, eliminating platform-specific design and fabrication. The fabrication of the laser is decoupled from the target platform by prefabricating its gain and cavity sections on separate substrates and transferring them onto the target platform. A platform agnostic design ensures consistent performance across platforms, facilitating the miniaturization of a diverse array of devices available on different platforms. The figure depicts some examples: fully integrated lithium niobate electro-optic combs, silicon nitride Kerr combs, and lithium tantalate modulators.}

\label{Fig1}
\end{figure}
Our goal is to develop a laser integration technique that is platform agnostic; a single design and process flow are used without modification for fabricating scalable integrated lasers across a large variety of platforms (see Fig. \ref{Fig1}). 
To realize this, we leverage micro-transfer printing, a heterogeneous integration technique allowing the simultaneous transfer of multiple fully processed devices from one chip to another, combining scalability with prefabrication\cite{margariti2023continuous,7999922}. We transfer a full laser in two printing steps and optimize the process to ensure near-identical laser performance on a variety of integrated photonic platforms without the need for additional design optimization or process development. 

We first show our technique's compatibility with a wide range of platforms, with refractive indices ranging from 1.7 to 2.5. We then experimentally validate our findings by showing near-identical laser characteristics for identically processed lasers integrated on TFLN and SiN platforms. Additionally, we demonstrate a laser-integrated resonant electro-optic (EO) comb by integrating our laser along with a frequency comb generator on the TFLN platform, resulting in a spectrum spanning 12 nm in the telecommunication region. The latter illustrates our platform-agnostic approach's ability to rapidly develop fully integrated devices on new platforms. 

\section{Results}
\subsection{Platform agnostic laser integration enabled by micro-transfer printing}
To guarantee that the laser behaves uniformly across platforms, its manufacturing process should not rely on the specific photonic platform being utilized and a single laser design and geometry should be universally applicable across platforms. Here, we first argue that micro-transfer printing is particularly suited for platform-agnostic fabrication since it enables processing the laser components separately and only then transferring them to the target platform. We then describe a platform agnostic design realized with a two-step transfer printing process: printing first the laser's cavity followed by its gain section. Finally, we describe how to use this process to fabricate platform-agnostic lasers over a large wavelength range, by combining different gain sections with variations of our optimized cavity.
 \\\\
 Micro-transfer printing \cite{mcphillimy2018high} \cite{TPreview} is a novel integration technique that combines the minimal post-processing of hybrid integration with the scalability of heterogeneous integration. Pieces (coupons) of the source material are patterned and underetched, then picked up through adhesion with a Polydimethylsiloxane (PDMS) stamp and finally printed on the target material. Due to the viscoelastic nature of PDMS, its adhesion is sensitive to pressure. Thus, by gradually releasing the stamp after printing, the coupon remains attached to the target. This controllable stamp-coupon adhesion enables printing on any material through Van der Waals (VdW) forces or a thin adhesion agent like benzocyclobutene (BCB).
 Importantly, this technique facilitates the simultaneous printing of multiple fully-processed coupons (utilizing arrayed stamps), thereby enabling large scale integration and device post-selection. 
The possibility of pre-processing complete laser stacks coupled with the non-specificity of the transfer technique makes it an ideal candidate for platform agnostic laser integration. 
 
 Previously, however, laser integration with micro-transfer printing required platform-specific considerations. It usually involves printing a III-V amplifier coupon onto a low-loss cavity on the target chip to facilitate lasing. The cavity can be an on-chip filter like an add-drop Vernier combined with a Sagnac reflector \cite{de2021iii}, or a quarter-wave shifted distributed feedback (QWSDFB) grating cavity \cite{Haq:20,Zhang:18}. In either case, platform-specific designs for the cavity are required, as the cavity resonance wavelength and shape depend on the platform index (defined as the refractive index of the waveguiding layer), the material stack, and the waveguide propagation losses.

\begin{figure}[ht!]
\centering\includegraphics[width=\textwidth]{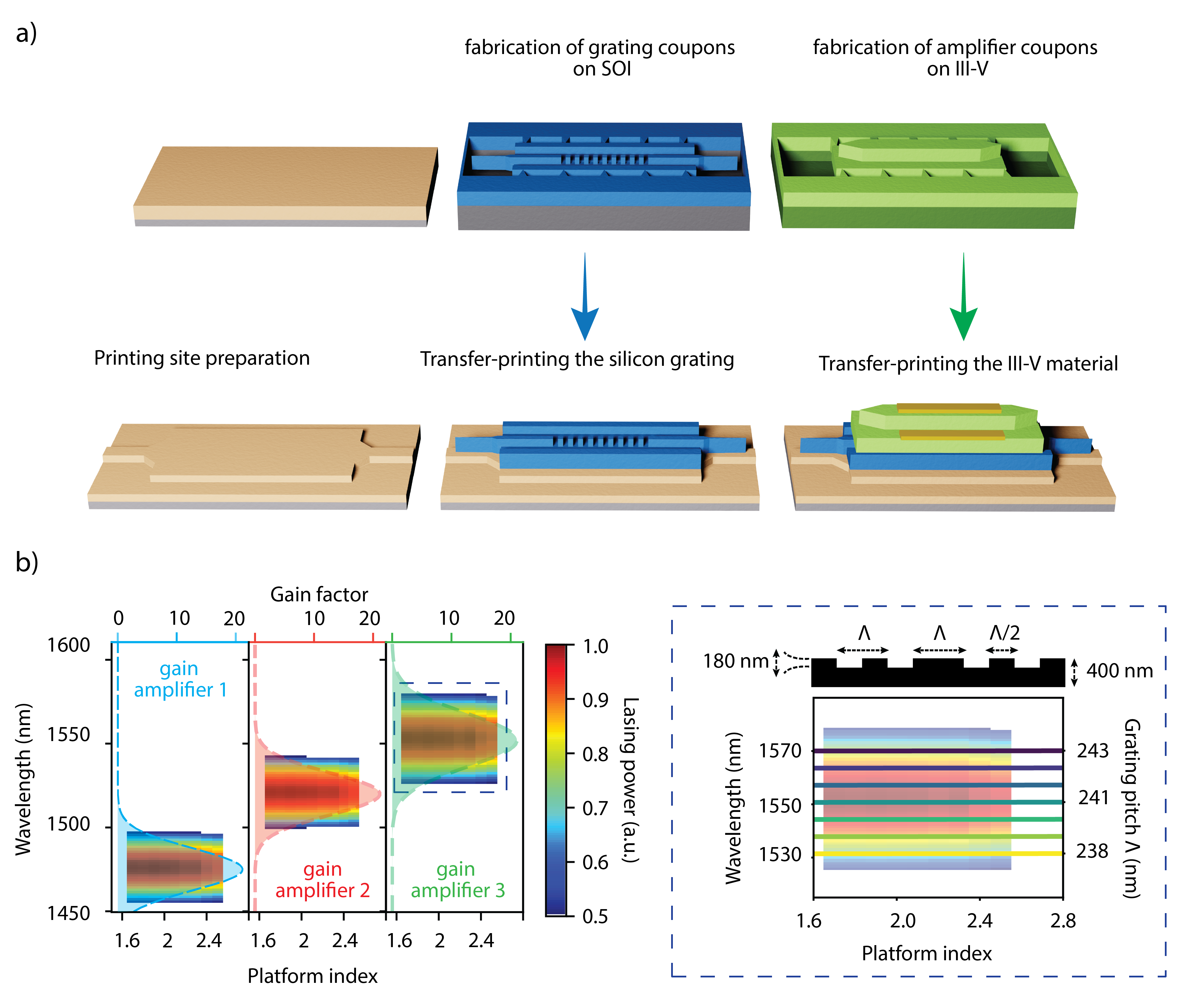}
\caption{\textbf{Fabrication process and design space for the platform-agnostic laser integration technique:} \textbf{(a)} The different fabrication steps of our technique are shown. Left panel: the printing site for the coupons is etched on the target photonics platform. Middle panel: the silicon coupon with a shallow-etched grating cavity and tapers is transferred from a prefabricated silicon-on-insulator wafer and printed onto the printing site. Right panel: a prefabricated III-V coupon with tapers included is then printed on top of the silicon grating serving as the gain section. This step is followed by the fabrication of electrical contacts to realize the final laser. \textbf{(b)}
The normalized lasing powers of lasers consisting of combinations of different silicon grating coupons and III-V gain coupons calculated for a range of platform indices and lasing wavelengths. Different gain bandwidths are provided by three III-V coupons with distinct stacks while the exact wavelength for each laser is fixed by selecting a silicon coupon with the corresponding grating pitch. Lasers of wavelengths between 1460 and 1580 nm can be fabricated on any platform with an index between 1.7 and 2.5 with the same set of prefabricated coupons. Relative lasing powers lower than 0.5 are not shown. The inset shows the lasing range for amplifier 3 with the cavity grating pitches and their corresponding lasing wavelengths superposed. A 1 nm shift in the pitch $\Lambda$ results in a 6.4 nm shift in the lasing wavelength. The wavelengths in between can be addressed by current tuning and thermal tuning. The cavity grating cross-section is shown on top of the inset. 
}
\label{Fig2}
\end{figure}

Furthermore, good coupling between the gain material and low index platforms is challenging since even narrow tapers in the usually ``thick'' ($\sim$ 4 \textmu m), high index (n $\approx 3.5$) III-V stack are not sufficient to ensure high-efficiency coupling to low index materials such as LN ($n=2.2$). This can be addressed with a thin intermediate layer of a higher index material such as Si (n$=3.4$) with narrow tapers fabricated using mature processing \cite{de2021iii}. Depositing this intermediate layer, however, requires extra in-situ processing steps such as deposition and etching which can compromise existing devices on the same chip. 

In our approach, we instead print a prefabricated silicon coupon on the platform material which combines the intermediate coupling section with a laser cavity, by including both a linear taper and a QWSDFB grating on the coupon (details in the next section). To finalize the laser we then print a III-V gain coupon on top of the silicon grating creating a Si/III-V DFB laser cavity (see Fig. \ref{Fig2}(a)).

This allows us to separately pre-process the III-V coupon containing the gain section and the silicon coupon containing both the cavity and the coupling section. It also opens up the possibility of combining a III-V gain section with different silicon cavities with distinct resonance wavelengths, generating a family of platform-agnostic lasers across the whole gain bandwidth. Importantly these silicon coupons with different cavities can be processed on the same source wafer. Furthermore, the wavelength range can be further extended by utilizing multiple III-V coupons, prefabricating them on different chips with distinct material stacks.

To show the potential of this approach we plot the calculated relative lasing powers and lasing wavelengths for different combinations of coupons in Fig. \ref{Fig2} (b) (details in Methods). We combine three existing prefabricated amplifier coupons (details in Supplementary section \ref{sec:amp_stack}) with optimized silicon coupons (see next section) with a fixed taper and varying QWSDFB pitches, where for each wavelength the matching QWSDFB pitch is assumed (see Fig. \ref{Fig2}(b) inset). The figure shows this approach can be used to fabricate efficient lasers for platforms with platform indices between 1.7 and 2.5 and a wide range of lasing wavelengths (1450$<\lambda<$1600 nm).

\subsection{Optimization of the coupon designs}
To optimize the platform-agnostic operation of the laser we have to address its three sections: the combined III-V/Si DFB laser cavity, the coupling section from the cavity to the silicon waveguide, and the coupling section from the silicon waveguide to the platform. We will show that the relatively high indices of the Si and III-V enable a platform-agnostic cavity and coupling section to the silicon waveguide by having strong optical confinement in the printed layers instead of the platform material. The silicon taper design will however be crucial to ensure efficient coupling between different platforms and the silicon waveguide.
We will also show that the wavelength independence is a serendipitous byproduct of the platform-agnostic optimization.

We optimize the different laser sections using Lumerical Mode and Eigenmode Expansion (EME) simulations with a model material stack for the considered photonic platforms consisting of a 600-nm thick platform material ($n$ = $n$\textsubscript{platform}) with partially etched waveguides 300 nm deep on top of 2-\textmu m of buried oxide ($n=1.55$) with a Si ($n=3.4$) substrate. We note however that the results obtained in this section are not dependent on the specific stack but are representative for any common material stack as will be shown in the experiments and Supplementary section \ref{sec:stack_comparison}.

\begin{figure}[ht!]
\centering\includegraphics[width=\textwidth]{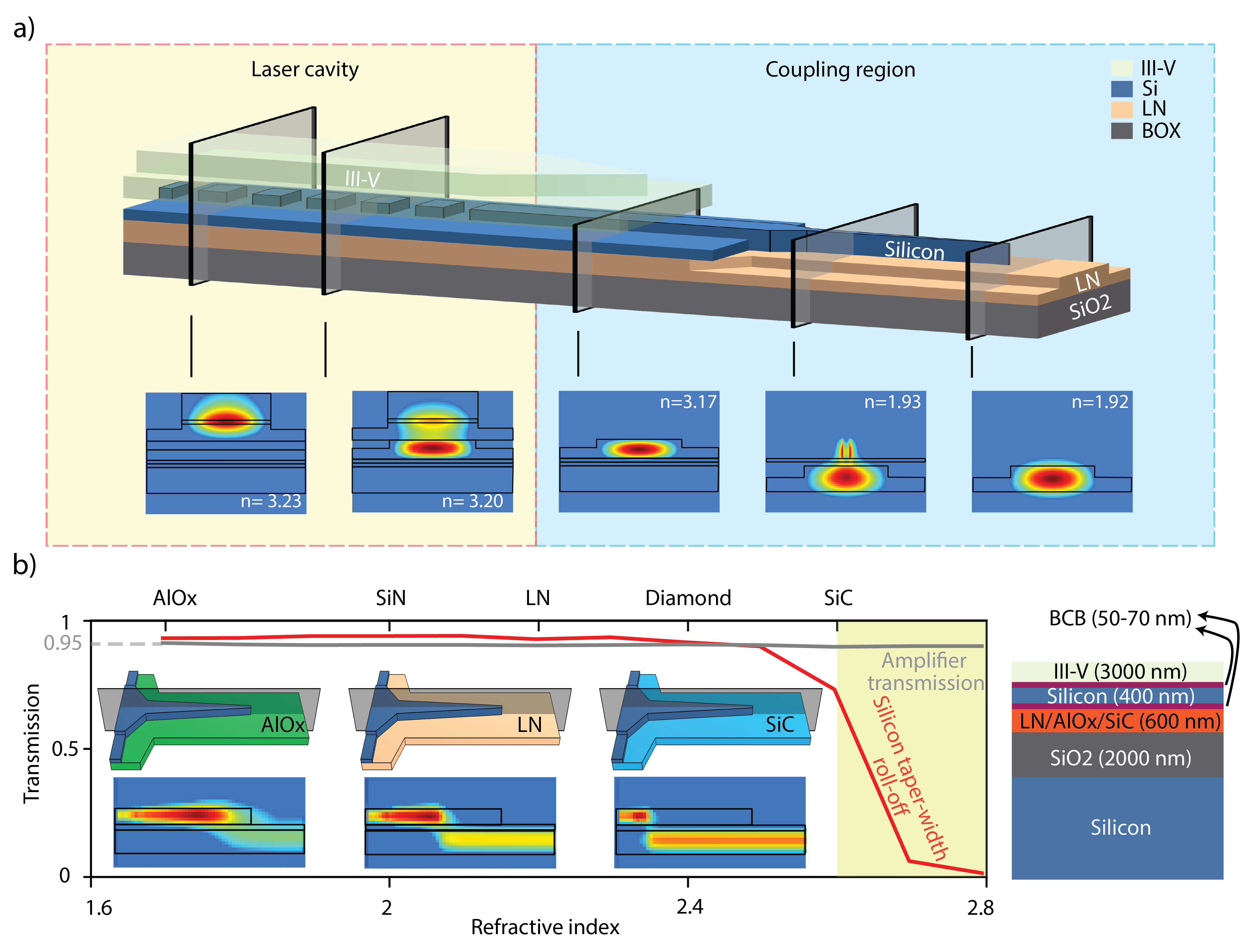}
\caption{\textbf{Platform-agnostic heterogeneous laser integration simulation and design:} \textbf{(a)} 3-D schematic of the different coupling interfaces between the platform material (indexes based on the thin film lithium niobate platform), the silicon coupon, and the amplifier coupon. The cross-sectional electrical field distribution of the TE optical modes at different locations and the associated effective index are shown below. In the laser cavity, the mode is solely confined in the silicon and III-V enabling the platform-agnostic lasing concept. \textbf{(b)} Simulated coupling efficiency (transmission) between the silicon coupon and the platform (red line) and the III-V and the silicon coupon (grey line) for various values of refractive indices. The transmission between the III-V and silicon is always 95\% since the mode does not interact with the platform material there. The insets are 3-D schematics, and Eigenmode Expansion (EME) simulation results (side-view) for the optical mode coupling between the Si coupon and AlO\textsubscript{x}, lithium niobate and SiC waveguides(600 nm thick device layer with a 300 nm slab on top of 2 \textmu m of oxide, 3 \textmu m wide waveguides). They show that the phasematching point shifts to wider points of the taper for higher platform indices. The yellow-shaded region shows a fast roll-off in the platform to Si transmission for indices higher than 2.5 due to the Si taper not being wide enough to match the effective index between the taper and the wider, relatively high index waveguide underneath. The material stacks used in these simulations are shown on the right. Both plots are considered for a wavelength of 1570 nm.}
\label{Fig3}
\end{figure}

First, we consider the laser cavity, a QWSDFB grating in the silicon with the III-V gain material on top, where the resonant wavelength $\lambda_{DFB}$ and reflectivity $R$ are determined by:
\begin{align}
\lambda_{DFB}=4\Lambda\mathrm{\frac{n_{etched} n_{unetched}}{n_{etched}+ n_{unetched}}}\\
R= \Big [ \mathrm{\frac{n_{etched}\strut^{\frac{2L}{\Lambda}}-n_{unetched}\strut^{\frac{2L}{\Lambda}}}{n_{etched}\strut ^{\frac{2L}{\Lambda}}+n_{unetched}\strut^{\frac{2L}{\Lambda}}}}\Big ]^2 \label{DFB_form}
\end{align}
There are three critical parameters for the cavity design, the length of the grating ($L$), the pitch $\Lambda$ and the effective refractive indices of the etched ($n_{etched}$) and unetched ($n_{unetched}$) grating sections. 
These effective refractive indices $\mathrm{n_{unetched}}\approx3.20$ and $\mathrm{n_{etched}}\approx3.23$ can be retrieved from mode simulations on different cross sections (Fig. \ref{Fig3} (a)) and show little dependence on the platform material since the optical mode is almost completely confined in the Si and III-V. As a consequence, the lasing wavelength for a grating with a specific pitch shifts only $\sim$ 50 pm for platforms indices ranging from 1.7 to 2.5 (see Supplementary section \ref{sec:grating_wl_shift}). 
By keeping $L$ fixed (450 \textmu m) and varying $\Lambda$, platform agnostic lasers at different wavelengths can thus be demonstrated (see Fig.\ref{Fig2} (b) inset).

Then we consider the coupling section consisting of two parts, first the light coming out of the laser cavity is coupled from the III-V/Si waveguide into the silicon waveguide after which this light gets coupled to the platform waveguide (see Fig.\ref{Fig3}(a)). 
For the first part, the optical mode is still mainly confined in the silicon and the III-V coupons so the coupling is not dependent on the photonics platform and a transmission of 95 \% is achieved between the cavity and the silicon waveguide using an adiabatic coupler in the III-V as described in \cite{Haq:20amp}. 

The second coupler, however, interfaces directly with the target photonics platform. To ensure coupling
between a variety of nano-photonic platforms and the silicon coupon, the coupler’s dimensions must cover the required effective index matching points for a wide range of platform indices. As we want to keep the printing site geometry of the platform fixed (3\textmu m waveguide at the coupler) this index matching point will correspond to different widths of the silicon coupler for different platforms.
To cover a large range of widths while keeping the coupling adiabatic, we adopt a linear taper geometry for the platform-silicon couplers. 

The taper is optimized by using Lumerical Eigenmode Expansion to calculate the transmission and varying both the taper dimensions and the platform index $\mathrm{n_{platform}}$ and keeping the wavelength fixed. By comparing these transmissions we can choose an optimal design within the fabrication constraints covering a wide range of indices. The optimized silicon taper has a narrow tip of 120 nm which tapers out to 340 nm width over 200 \textmu m (see Supplementary section \ref{sec:geometry}) and has a simulated transmission at 1570 nm of $ > 93\% $ for $n$\textsubscript{platform}=1.7-2.5 , with a peak transmission of 99.5\% for $\mathrm{n_{platform}} = 2$ (see Fig. \ref{Fig3}(b)). 
For different material indices the optical mode transfers from the silicon taper to the platform at different spots along the taper, this is where there is a match between their effective refractive indices. When the platform index rises, the transition point moves in the direction of the wider section of the taper. This shift continues until a substantial decline in coupling efficiency is observed for platform indices exceeding 2.5, primarily due to the constrained width (340 nm) of the wider section of the taper (Fig. \ref{Fig3} (b)-insets).

The width of the silicon taper is constrained there because it must be significantly smaller than the width of the printing site underneath (3 \textmu m) to avoid a dramatic drop in the coupling efficiency in case of lateral misalignment. For the optimized taper and $n$\textsubscript{platform}=1.7-2.5, $\lambda$=1570 this drop is less than 5\% for a lateral offset of 500 nm of the silicon coupon, which is 3 times the standard deviation($(3\sigma= \pm$ 500 nm) of state of the art transfer printing tools (see Supplementary section \ref{sec:tp_offset} for the wavelength dependence). For higher index platforms, one could use a wider printing site but this would come at the cost of exciting higher order modes in the waveguide for lower index platforms. 

Also a lateral offset of 500 nm for III-V coupon results in less than 2 \% drop in transmission across our index range at 1570 nm (see Supplementary section \ref{sec:tp_offset} for the wavelength dependence).

We additionally note that the transmission of the full coupling section consisting of the silicon and III-V tapers is not significantly affected by the operation wavelength, with $T (\lambda, \mathrm{n_{platform}}) > 80\% $ for $n$\textsubscript{platform}=1.7-2.5 and $\lambda$\textsubscript{laser}=1460-1580 nm spanning most of the gain bandwidth of the three amplifiers considered (see Supplementary section \ref{sec:stack_comparison}). This wavelength insensitivity was not specifically optimized for but is a byproduct of the platform agnostic design. While we use a single taper design here, the operating range could further be extended by using multiple designs for different wavelength and index ranges.

\subsection{Experimental demonstration of platform agnostic laser integration}
\begin{figure}[ht!]
\centering\includegraphics[width=\textwidth]{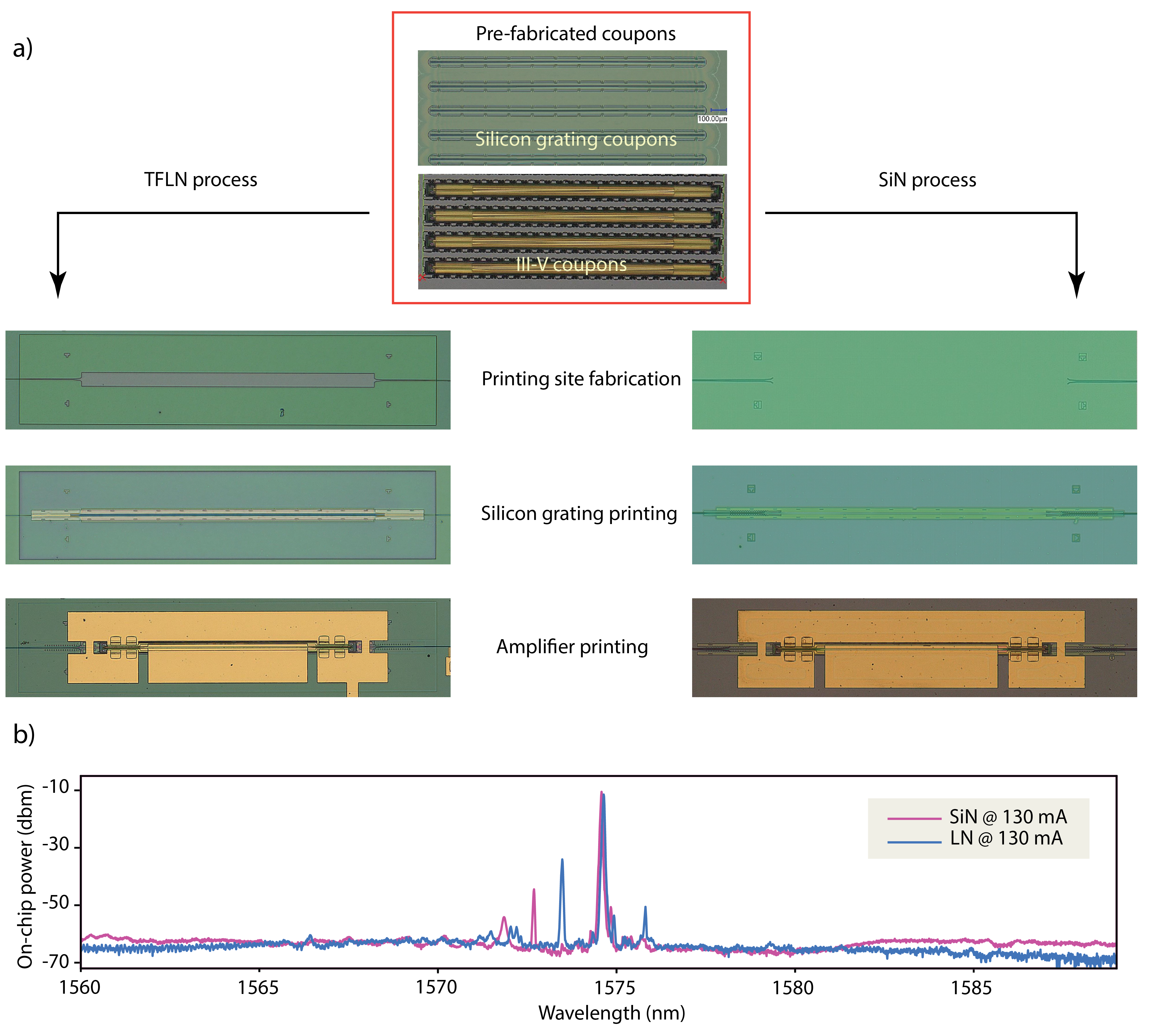}
\caption{\textbf{Experimental results for platform-agnostic lasers transfer printed onto thin-film lithium niobate and silicon nitride platforms.} \textbf{(a)} The silicon cavity coupon, and III-V gain coupons (shown at the top) are transferred to printing sites in TFLN and SiN. The laser fabrication is finalized by patterning the extended electrodes for driving the lasers. \textbf{(b)} Laser emission spectrum for both lasers at 130 mA showing a near identical emission wavelength at $\sim$ 1574.5 nm. The limited sideband suppression is due to the minimum resolvable features in the silicon grating.}
\label{Fig4}
\end{figure}

To experimentally demonstrate the platform-agnostic laser integration technique we fabricate lasers using the same coupons on SiN and LN. We first describe the fabrication of the Si and III-V coupon chips, and then we elaborate on the preparation of the SiN and TFLN platform chips, followed by the printing process. Finally, we discuss the laser measurements on both platforms and show that the laser performance is nearly identical. 

For the Si coupons we start from a silicon-on-insulator sample with a 400 nm silicon layer on top of a 2 \textmu m BOX layer. We partially etch (180 nm deep) the silicon waveguides and the QWSDFB grating ($\Lambda=247$ nm) and define the coupons by fully etching a rectangular trench around the waveguides with grating, while retaining some tethers connecting the coupon to the surrounding silicon (see Methods). The coupons are then underetched, resulting in suspended coupons with partially-etched waveguides and a QWSDFB grating (details in Methods).
The processing of the III-V amplifier coupons is described in \cite{Haq:20amp}. In fact the amplifier coupons we use come from the same chip used in \cite{Haq:20amp}, which shows the robustness of these coupons and makes it very simple to integrate them on new platforms as they do not need to be refabricated. 
The prefabricated silicon and III-V source chips are then ready to be used both containing hundreds of coupons to be printed. 

The SiN target chip is fabricated on a 300 nm thick SiN on 3.3 \textmu m of oxide on a silicon substrate with fully-etched waveguides while the TFLN target chip is fabricated on 600 nm of x-cut TFLN with partially etched (300 nm etch depth) waveguides on 2 \textmu m oxide on a silicon substrate \cite{zhang2017monolithic,shamsreduced2022} and an 800-nm thick oxide cladding (details in Methods). 

The printing site for both SiN and TFLN is defined as a single-mode waveguide tapering to a 3 \textmu m wide section which is 130 \textmu m long after which it tapers to $>60$ \textmu m wide section. The 3 \textmu m waveguide part will be where the Si taper couples the light to the platform, while the $> 60$\textmu m wide section is where the cavity will be printed (see Fig. \ref{Fig2} (a)).
For the LN, we then open a window on top of the printing site, etching the cladding down to 100 nm to ensure good coupling efficiency between the coupons and the LN, while the SiN is already uncladded.

After the circuits on the LN and SiN target chips are defined, the fabrication steps are identical. We first spin coat 50-70 nm of Benzocyclobutene (BCB) to provide extra adhesion for the transfer printing process and deposit an Al\textsubscript{2}O\textsubscript{3} etch stop layer (35 nm). We then transfer-print identical silicon coupons containing the QWSDFB ($\Lambda$ = 247 nm) and an access waveguide on top of the printing site on each platform. Subsequently, the silicon tapers are patterned and fully etched with the Al\textsubscript{2}O\textsubscript{3} etch stop layer protecting the underlying platform. This step requires a degree of process compatibility as the patterning of the taper can be subtly influenced by the underlying platform. As part of future improvements, it will be incorporated into the coupon preprocessing, mitigating any platform-specific fabrication considerations.

This is followed by another round of BCB spin coating and printing the same amplifier (amplifier 3) coupons with included tapers on both chips. We then cover everything with 3 \textmu m of BCB, open the vias, pattern the electrodes, and deposit the Ti-Au for the final contacts (details in Methods).
The amplifier coupons printed have gain centered around 1550 nm and the QWSDFB grating ($\Lambda$ = 247 nm) in the silicon coupons will select the target wavelength within the gain bandwidth (see Fig. \ref{Fig2}(b) inset). 

We use a current source to pump the lasers and measure a threshold current of $\sim$ 75 mA for both devices (see Supplementary section \ref{sec:tune lasers}). At 130 mA the lasers reach their maximum lasing power as higher drive current will result in lower power due to self-heating effects. At this drive current, the emission spectra of the laser on SiN and LN are near-identical (see Fig. \ref{Fig4}(b) ): single-mode lasing with a lasing wavelength of $\sim$ 1574.5 nm and an on-chip power of -10.5 and -11.5 dBm respectively. At different drive currents, the lasing powers and wavelengths also remain very similar (see Supplementary section \ref{sec:compare lasers}).

It is important to note that the Si and III-V coupons printed are identical not only in design but that they come from the same prefabricated chip. This means any systematic errors introduced during fabrication will affect both coupons ensuring identical operation and allowing for pretesting on a test platform e.g. SiN before printing them on a harder-to-fabricate platform like low-loss LN. 
 
Aside from systematic fabrication errors, random local imperfections are minimal for both silicon and III-V due to their process maturity. Moreover, the lasing wavelength can be finely tuned by changing the pump current (see Supplementary section \ref{sec:tune lasers} ), further reducing the impact of random local fabrication errors. 
Finally we remark that the compared lasers are printed on platforms with different platform indices as well as different material stacks: 300 nm SiN on 3300 nm SiO\textsubscript{2} on Si and 600 nm LN on 2000 nm SiO\textsubscript{2} on Si,while retaining near-identical laser performance, demonstrating that the process is truly platform agnostic (see also Supplementary section \ref{sec:stack_comparison}).

\subsection{Fully integrated electro-optic frequency comb} 
\begin{figure}[!htb]
\centering\includegraphics[width=0.5\textwidth]{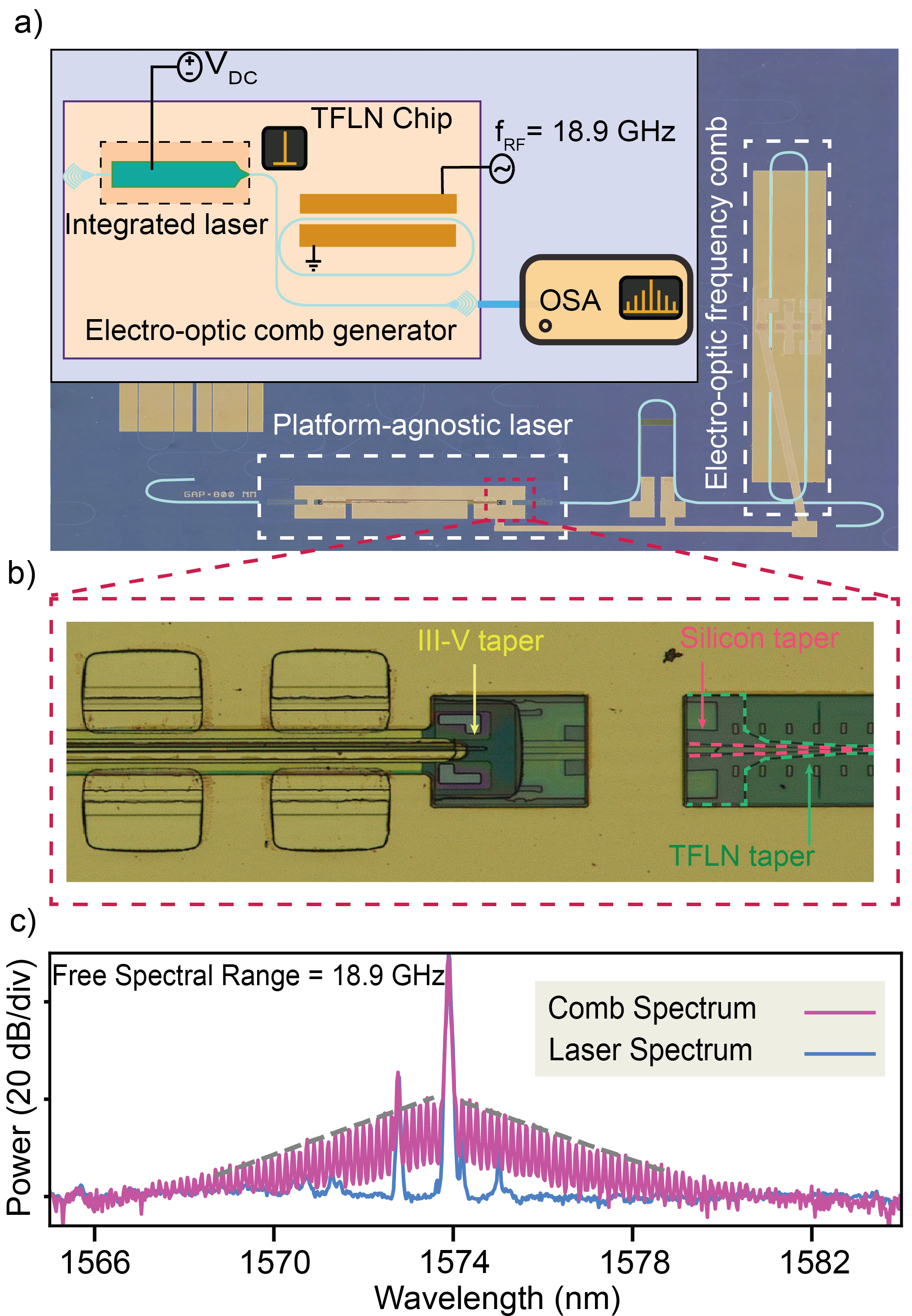}
\caption{\textbf{Experimental demonstration of laser-integrated electro-optic frequency comb source.}\textbf{(a)} Color-corrected micrograph of the fabricated resonant electro-optic frequency comb source integrated with a platform agnostic laser. the inset shows the schematic of the measurement setup.\textbf{(b)} Zoom-in image of the coupling region showing all transitions from TFLN to the silicon coupon, and from the silicon to the III-V amplifier coupon. \textbf{(c)} Measured output spectrum of the EO comb generated from the microring resonator, demonstrating a bandwidth exceeding 12 nm and more than 80 comb lines with a spacing of 18.9 GHz and a slope of $-3.5$ dB/nm.}
\label{Fig5}
\end{figure}
To demonstrate the utility of platform-agnostic laser integration we use the laser on LN to realize a fully integrated electro-optic comb. We first discuss why this is a relevant example, then we detail the co-integration of the laser with the EO comb generator and discuss the characteristics of the resulting comb.

While numerous examples of on-chip photonic advancements can benefit from laser co-integration, we specifically focus on using our technique to integrate a laser with an electro-optic frequency comb generator. Compared to other frequency comb sources, EO combs have several advantages such as fundamental coherence, smaller comb spacing, flexibility in the center operating wavelength, and reconfigurability of the comb spacing \cite{zhang2019broadband}. Despite earlier demonstrations of laser integration with thin-film lithium niobate platforms \cite{shams2022electrically,OpdeBeeck:21,zhang2023heterogeneous}, there is yet to be a demonstration of a fully integrated resonant-based electro-optic comb source. 

We realize this device by first fabricating a TFLN circuit consisting of a transfer printing site connected through single-mode waveguides to a high Q racetrack cavity with electrodes along the straight arms (Fig. \ref{Fig5}(a)) and then printing the laser on the transfer printing site as described above. 

An EO comb can then be generated by coupling the laser into the racetrack resonance by tuning its frequency to a resonance frequency $\mathrm{f_0}$. The electro-optic effect of LN is then used to modulate the refractive index of the two arms of the racetrack resonator, resulting in pure phase modulation of the circulating light. By matching the modulation frequency to the ring's free spectral range (FSR) frequency $\mathrm{f_{RF}=f_{FSR}}$, two sidebands are generated at $\mathrm{f_0-f_{FSR}}$ and $\mathrm{f_0+f_{FSR}}$ which in turn will be resonant, build up in power, and generate further sidebands, building up a comb with a spacing of $\mathrm{f_{FSR}}$. The efficiency of generating sidebands is proportional to $e^{-\frac{a}{QV}}$ \cite{zhang2019broadband}, with $a$ the round trip loss, $Q$ the quality factor of the racetrack resonator (here $Q\sim500k$) and $V$ the RF voltage applied to the resonator arms \cite{zhang2019broadband}.

Here we tune the lasing wavelength into the racetrack resonance at 1574 nm using current tuning ($1.5$ nm tuning range: see Supplementary section \ref{sec:tune lasers}) by driving it with 108 mA (2V). At this operation point the integrated laser has -14 dBm of output power and a linewidth of 40 MHz. We then apply EO-modulation at the FSR frequency of 18.9 GHz using an RF source followed by a power amplifier, amplifying the signal to $P_{\mathrm{RF}} =$ 30 dBm, although most of this power is reflected as the device is not impedance matched. The generated spectrum consists of ~80 comb lines with a bandwidth of more than 12 nm and a slope of $-3.5 \mathrm{\frac{dB}{nm}}$. (see Fig. \ref{Fig5}(c)). Although this is a relatively small bandwidth, it could be broadened considerably by incorporating a secondary ring resonator as in \cite{hu2022high}, increasing the laser power or improving the quality factor of the racetrack resonator. Our integration technique could also be used to combine EO combs centered around different wavelengths on the same chip by using prefabricated silicon coupons with different QWSDFB grating pitches.

The fully integrated comb shown here is a critical step towards the realization of compact frequency-agile spectrometers \cite{shams2022thin} with high efficiency \cite{hu2022high}. It exemplifies the rapid integration that is possible with the platform-agnostic laser integration technique as the lasers can be pretested on a test platform while providing design flexibility and limited in-situ tunability of the wavelength.

\section{Discussion}
In summary, we have introduced a platform-agnostic technique for rapidly integrating single-mode lasers with center wavelengths between 1460 and 1580 nm onto photonics platforms with refractive indices ranging from 1.7 to 2.5. This technique consists of a two-step transfer printing process, printing first a silicon coupon containing the cavity and the coupling section and then printing a III-V coupon containing the gain section. The design of the coupling section and the strong confinement of the light in the printed layers avoid the need for any platform-specific design or fabrication.
This enables off-chip fabrication and pretesting and optimization of the laser on a test platform ensuring predictable and high-yield lasing operation for future platforms. Importantly the silicon and III-V coupons can be prefabricated and remain stable for years, allowing for off-the-shelf processing. Moreover, different cavity and gain coupons can be readily combined allowing for flexibility in lasing wavelength and other laser characteristics.
This approach is particularly valuable for prototyping and rapidly integrating lasers on evolving platforms \cite{Dorche:23, lu2018aluminum} and co-integrating lasers of different wavelengths \cite{chauhan2021visible,franken2021hybrid,tran2022extending}.

We have validated our approach by successfully implementing lasers on both SiN and TFLN platforms, achieving nearly identical performance in terms of lasing wavelength and power. Furthermore, we have demonstrated the utility and ease of integration of our technique by showcasing a fully-integrated resonant electro-optic frequency comb on the TFLN platform, generating more than 80 comb lines over 12 nm from the on-chip laser. 

Looking ahead, our focus will be on transferring the remaining on-chip post-processing to preprocessing by etching the tapers of the silicon coupon before printing instead of after. This would simplify on-chip processing to just the two transfer printing steps and depositing electrodes, avoiding post-processing of the final device.

Aside from optimizing the technique we aim to integrate lasers on various platforms to realize fully integrated devices with different applications. The implementation of our miniaturized, power-efficient integrated lasers with various comb sources would pave the way for compact, high-performance transceivers and microwave photonic systems. \cite{marin2017microresonator, corcoran2020ultra, hu2018single,spencer2018optical}. 
Similarly, the next generation of compact photonic processors \cite{zhou2021large,bai2023microcomb, bai2023photonic} such as optical neural networks \cite{shen2017deep}, can benefit from our platform agnostic approach to integrate a large number of laser sources with different characteristics on the same chip.
Finally, further optimization of the III-V gain and silicon cavity coupons would improve the laser linewidth and output power, enabling demanding applications in metrology.

\section{Materials and methods}
\subsection{Calculation of lasing powers for different coupon combinations}
In Fig \ref{Fig2} (b) the relative lasing power coupled to the platform is calculated with:
\begin{align}
P_{\mathrm{laser}}(\lambda, \mathrm{n_{platform}})= T (\lambda , \mathrm{n_{platform}}) \times \frac{\alpha_m(\lambda, \mathrm{n_{platform}}) }{\alpha_m(\lambda, \mathrm{n_{platform}})+\alpha_i}\frac{h\nu}{q} \Big[ I-I_{\mathrm{th}}(\lambda, \mathrm{n_{platform}}) \Big]
\label{laser_formula}
\end{align}
Here $T (\lambda , \mathrm{n_{platform}})$ is the power transmission from the laser to the platform material, largely determined by the taper in the silicon coupons , $\alpha_i$ is the internal cavity loss, $h\nu$ is the photon energy and $q$ is the electron charge. $\alpha_m$ is the distributed mirror loss, defined as $\alpha_m=\frac{1}{L}\ln{\frac{1}{R}}$ with $R(\lambda,\mathrm{n_{platform}})$ the simulated QWSDFB grating reflection dependent on both the wavelength and platform index and $L$ the cavity length. $I$ is the drive current and $I_{th}$ is the threshold current which can be derived for the different amplifiers from the measured wavelength dependent small signal gain $g(I,\lambda)$ and the QWSDFB reflection $R(\lambda,\mathrm{n_{platform}})$. This formula can be interpreted as the amount of surplus electrons converted to photons above threshold $\frac{h\nu}{q} \frac{\alpha_m }{\alpha_m+\alpha_i}( I-I_{\mathrm{th}} )$ multiplied by the fraction that is coupled out of the cavity $\frac{\alpha_m }{\alpha_m+\alpha_i}$ and the coupling from the laser to the platform $T (\lambda , \mathrm{n_{platform}})$ . The powers shown are calculated assuming $I=120mA$ and using: the simulated silicon grating reflection, assuming the correct pitch at each wavelength, the extrapolated (for amplifiers 1, and 2) and measured (for amplifier 3) gain spectra, and the simulated taper transmission (More details in supplementary section \ref{sec:formula1_details}).

\subsection{ Fabrication}

For the Si coupons we start from a silicion-on-insulator sample with a 400 nm silicon layer on top of a 2 \textmu m BOX and a Si substrate. We then partially etch trenches defined by e-beam lithography with an etch depth of 180 nm to define the silicon waveguides and the QWSDFB grating ($\Lambda$=247 nm) using a reactive ion etch (RIE) with CF\textsubscript{4}, H\textsubscript{2} and SF\textsubscript{6}. 
To define the coupons we use the same RIE recipe to fully etch a rectangular trench around the waveguides with grating, while retaining some tethers connecting the rectangular coupon to the surrounding silicon.
Then we underetch this silicon coupon by etching away the silicon oxide underneath using a vapor HF etch. At this point we have suspended coupons with on top partially etched waveguides with a QWSDFB grating and this source chip is ready to be used. After the coupons are printed tapers are fully ethed using the same RIE recipe.

The processing of the III-V amplifier coupons is described in \cite{Haq:20amp} . In fact the amplifier coupons we use come from the same chip used in \cite{Haq:20}, which shows the robustness of these coupons and makes it very simple to integrate them on new platforms as they do not need to be refabricated. 
These prefabricated coupons are then ready to be printed. 

The SiN target chip is fabricated on a 300 nm thick LPCVD grown SiN material on 3.3 \textmu m of thermally grown oxide. The fully-etched SiN waveguides are defined using e-beam lithography (positive tone AR-P 6200.13 ) followed by an RIE with CF\textsubscript{4}, and H\textsubscript{2} chemistry.

The TFLN target chip is fabricated on 600 nm of x-cut TFLN on 2 \textmu m of thermally grown oxide on a silicon substrate. The waveguides are patterned using E-beam lithography with a negative tone resist (HSQ) and partially etched (300 nm etch depth) with RIE (Ar+ ) and are later cladded with 800 nm of Inductively Coupled Plasma Chemical Vapour Deposition (ICPCVD) SiO2, followed by an annealing step to recover damages caused by ion implantation \cite{zhang2017monolithic,shamsreduced2022}. The 300 nm etch depth is chosen to maintain the optical confinement, while providing moderate overlap between the optical and microwave fields. Light is coupled from the laser to the race-track resonators (1.6 µm wide waveguides to reduce losses) using a symmetric coupler in the TFLN to reject higher order modes.

\subsection{Measurement conditions}
In all experiments the lasers are driven using a Keithley 2400 as the current source. The laser output is measured from one of the grating couplers as light is emitted to both sides, the listed powers are all single-side output. The comb and laser spectra are acquired using an Anritsu MS9740A Optical spectrum analyser. The RF voltage applied to the racetrack resonator in the comb experiment is generated by the Rohde \& Schwarz SMR40 Signal Generator. All measurements are done with a temperature controller set to ${25}^\circ$C. 
\section{Data availability}
The datasets generated during and/or analysed during the current study are available from the corresponding author on reasonable request.

\section{Acknowledgments}
This work was funded by Defense Advanced Research Projects Agency (DARPALUMOS) (HR0011-20-C-0137) , Air Force Office of Scientific Research (AFOSR) (FA9550-19-1-0376) and the European Research Council (ERC) starting grant ELECTRIC (759483). 
LN device fabrication was performed at the Center for Nanoscale Systems (CNS), a member of the National Nanotechnology Coordinated Infrastructure Network (NNCI), which is supported by the National Science Foundation under NSF Grant No. 1541959.

\section{Conflict of interest}
M.L. is involved, as a co-founder and board member of HyperLight Corporation, in developing lithium niobate technologies. The remaining authors declare no competing interests.

\section{Author contributions}
 I.L.L and B.K developed the concept of platform agnostic laser integration. A.S.A and I.L.L conceived the experiment, and designed the devices. I.L.L fabricated and designed the lasers. A.S.A fabricated the TFLN chips with help from D.R. I.L.L performed the laser measurements. A.S.A characterized the passive TFLN devices. I.L.L fabricated SiN devices and performed the laser transfer printing on both TFLN and SiN devices with help from C.O.B, S.C, S.P. MB fabricated, printed and characterized amplifiers 1 and 2 on a SiN platform. I.L.L performed the electro-optic comb measurements with feedback from A.S.A. A.S.A, and I.L.L wrote the manuscript with help from other co-authors. G.R helped with the overall project. M.L and B.K supervised the project.

\newpage
\clearpage
\bibliographystyle{unsrt}
\bibliography{sample}
\newpage
\clearpage

\title{Supplementary information: Integrated resonant electro-optic comb enabled by platform-agnostic laser integration}
\date{}
\author{}
\maketitle
\setcounter{figure}{0}
\setcounter{section}{0}

\section{Characteristics of the amplifier coupons} \label{sec:amp_stack}
\begin{table}[!htb]
\centering
\caption{III-V SOA epitaxial layer stack for amplifier 3, taken from \cite{Haq:20amp} .}
\label{table:layer-stack}
\small
\begin{tabular}{|c|c|c|c|c|c|}
\hline
\textbf{Layer} & \textbf{Layer type} & \textbf{Material} & \textbf{Thickness [nm]} & \textbf{Doping level [cm\textsuperscript{-3}]} & \textbf{Dopant} \\
\hline
27 & Cap layer & InP & 100 & nid & \\
26 & Contact P & InGaAs (lattice matched) & 100 & $> 1 \times 10^{19}$ & Zn/C \\
25 & Contact P & InGaAs (lattice matched) & 100 & $\approx 1 \times 10^{19}$ & Zn \\
24 & Cladding P & InP & 1000 & $\approx 1 \times 10^{18}$ & Zn \\
23 & Cladding P & InP & 500 & $\approx 5 \times 10^{17}$ & Zn \\
22 & Etch stop & InGaAsP ($\lambda_g = 1.17 \mu$m) & 25 & $\approx 5 \times 10^{17}$ & Zn \\
21 & Transition & (Al$_{0.9}$Ga$_{0.1}$)$_{0.47}$In$_{0.53}$As & 40 & & \\
20 & SCH & (Al$_{0.7}$Ga$_{0.3}$)$_{0.47}$In$_{0.53}$As & 75 & & \\
14-19& Barrier x 6& (Al$_{0.45}$Ga$_{0.65}$)$_{0.51}$In$_{0.49}$As & 10 & & \\
8-13& Well x 6& (Al$_{0.25}$Ga$_{0.75}$)$_{0.3}$In$_{0.7}$As & 7.5 & & \\
7& Barrier & (Al$_{0.45}$Ga$_{0.65}$)$_{0.51}$In$_{0.49}$As & 6 & & \\
6& SCH & (Al$_{0.7}$Ga$_{0.3}$)$_{0.47}$In$_{0.53}$As & 75 & & \\
5& Transition & (Al$_{0.9}$Ga$_{0.1}$)$_{0.47}$In$_{0.53}$As & 40 & $1 \times 10^{18}$ & Si \\
4& Cladding N & InP & 200 & $2 \times 10^{18}$ & Si \\
3& Cladding N & InP & 60 & nid & \\
2& Sacrificial & InGaAs (lattice matched) & 50 & nid& \\
1& Sacrificial & AlInAs (lattice matched) & 500 & $\approx 0.5 \times 10^{18}$ & Si \\
0& Buffer layer & InP & 150 & & \\
\hline
\end{tabular}
\end{table}
In the main text we refer to a set of amplifier coupons which were pre-fabricated in our cleanroom. Table \ref{table:layer-stack} shows the amplifier stack for "amplifier 3" which we used in the experiments. The other amplifier stacks mentioned are very similar but with adjusted multi quantum wells to shift the center gain wavelength. 
\section{Details of the lasing power formula}\label{sec:formula1_details}
In the main text the following formula is used to calculate the lasing powers:
\begin{align}
P_{laser}(\lambda, n_{platform})= T (\lambda , n_{platform}) \times \frac{h\nu}{q} \frac{\alpha_m }{\alpha_m+\alpha_i}( I-I_{th} )
\label{laser_formula_current_supp}
\end{align}
Here $\alpha_m$ is the distributed mirror loss defined as $\alpha_m=\frac{1}{L}\ln{\frac{1}{R}}$ with $R(\lambda,n_{platform})$ the simulated QWSDFB grating reflection dependent on both the wavelength and platform index. The threshold current $I_{th}$ can be derived for the different amplifiers from the measured wavelength dependent small signal gain $g(I,\lambda)$ and the QWSDFB reflection $R(\lambda,n_{platform})$\cite{verdeyen1989laser} . 
ent. 

To measure the wavelength dependent gain factor ($g(I,\lambda)$) we first print modified silicon coupons on top of a SiN transfer printing site. These modified silicon coupons include tapers and a waveguide but omit the cavity to enable direct coupling from the amplifier coupon to the SiN waveguides. We print all three amplifier coupons onto the silicon test coupons. The silicon couplers, and the amplifier coupons in this experiment are identical to the lasing experiment. 
We then pump the final printed amplifiers with different currents ( $50-140 mA$) and extract the amplified simulated emission by monitoring the spectra on an optical spectrum analyzer. We then select coupon 3 to be used for the final lasing experiment, and measure its small signal amplification for different currents. This is done by sweeping a tunable laser (SANTEC TSL-510) and monitoring the spectrum and the peak power. We later extrapolate the gain factor for other coupons by assuming a similar wavelength dependence as the gain of coupon 3 but shifting its center wavelength based on the ASE center wavelength. The length ($L$) of the Si cavity length and the estimated reflection coefficient of the Si QWSDFB are 450 \textmu m and $R=0.95$, respectively. 

\section{Generalization of the material stack for the platform agnostic laser integration technique} \label{sec:stack_comparison}
In the main text we use a model material stack for the considered photonic platforms consisting of a 600 nm thick platform material with partially etched waveguides 300 nm deep on top of 2-$\mu$m of buried oxide ($n=1.55$) with a Si ($n=3.4$) substrate. Here we show however that the results generalize very well to other common material stacks. We first consider variations in the waveguide layer thickness and etch depth, then we address differences in the BOX thickness and finally we elaborate on the effects of having different substrates.

\begin{figure}[!htb]
\centering\includegraphics[scale=1]{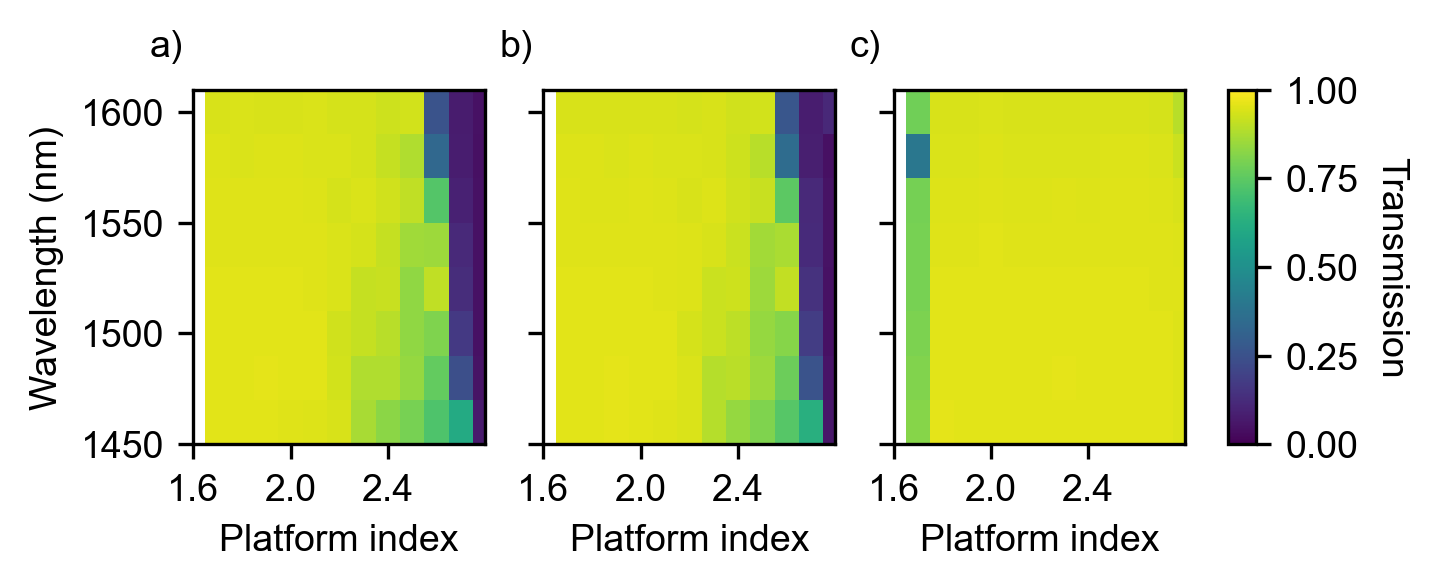}
\caption{\textbf{Power transmission from platform agnostic lasers to platforms with different material stacks} When considering different model stacks we see the double taper maintains a high transmission across different platform indices and wavelengths. The stacks considered all have a waveguide layer on top of 2 $\mu$m of BOX. The waveguide layers are a) a 600 nm rib waveguide with a 300 nm slab b) a 600 nm fully etched strip waveguide c) a 300 nm fully etched strip waveguide
}
\label{T_stacks_wl}
\end{figure}

\begin{figure}[!htb]
\centering\includegraphics[scale=1]{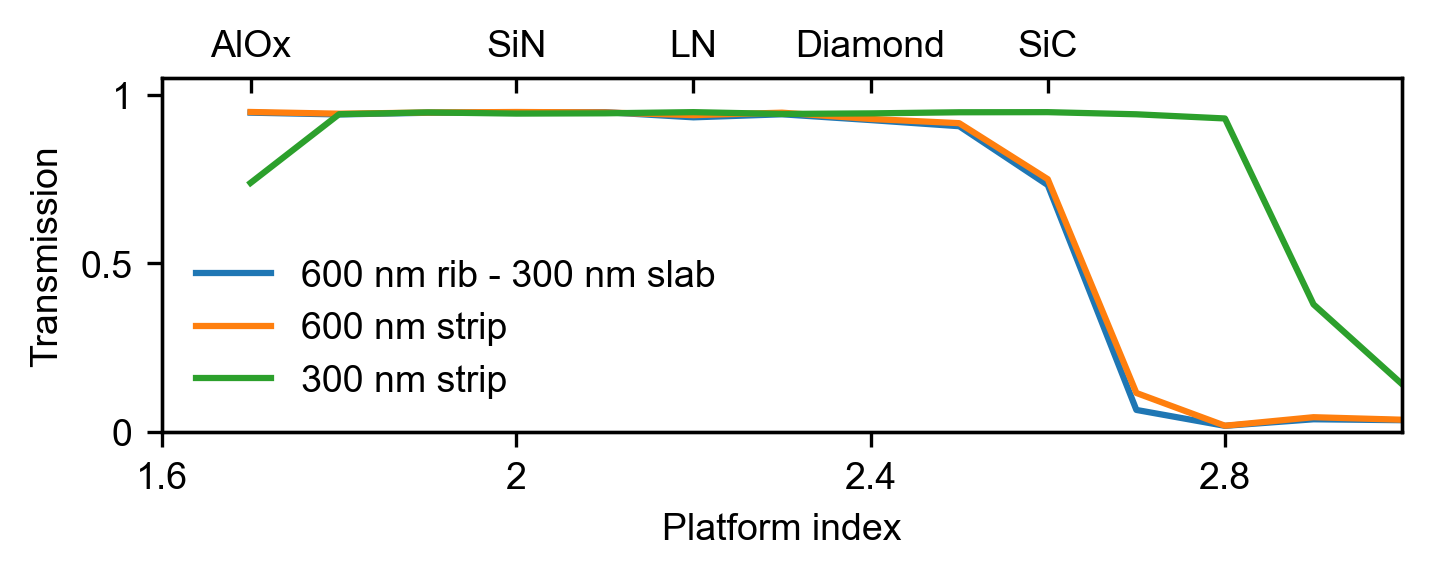}
\caption{\textbf{Power transmission from a platform agnostic laser at 1470 nm to platforms with different material stacks} When considering different model stacks we see the double taper maintains a high transmission across different platform indices at 1570 nm. The stacks considered all have a waveguide layer on top of 2 $\mu$m of BOX. The waveguide layers are a 600 nm rib waveguide with a 300 nm slab, a 600 nm fully etched strip waveguide and a 300 nm fully etched strip waveguide. The 300 nm stack shows an extended and shifted operation range of $n_{platform}$=1.8-2.8 
}
\label{T_stacks_1550}
\end{figure}

We consider 3 waveguide configurations: a 600 nm wire, a 300 nm wire and a 600 nm rib waveguide with a 300 nm slab as considered in the main text. 
As shown in the main text the waveguide layer does not affect the cavity and the coupling from the cavity to the silicon waveguide since the mode is almost completely confined in the silicon and the III-V. Only the coupling section from the silicon waveguide to the platform material can thus be affected. In Fig. \ref{T_stacks_wl} and \ref{T_stacks_1550} we consider the full coupling from the laser to the platform though, to show the full picture.Fig. \ref{T_stacks_wl} shows that when varying wavelengths and platform indices the dual taper transmission is similar for the different material stacks. Focusing on the transmission at 1570 nm we see in Fig. \ref{T_stacks_1550} that the transmission is larger than 90 \% , 91\% ad 74\% for $n_{platform}=1.7-2.5$ for the 600 nm rib 300 slab, 600 nm strip and 300 nm strip respectively. Although for the 300 nm strip the operation platform indices are just shifted to $n_{platform}=1.8-2.7$ where transmission is higher than 94\%.This result is also borne out in the experiment in the main text as the laser performance is near-identical and the stacks of SiN and LN platforms used in the experiment are different: 300 nm wire on $3.3 \mu m$ SiO\textsubscript{2} on a silicon substrate and 600 nm rib on $2 \mu m$ SiO\textsubscript{2} with a 300 nm slab on a silicon substrate respectively. 
Looking at different wavelengths from 1460 nm to 1580 nm for the operating ranges $n_{platform}=1.7-2.5$ for the for the 600 nm rib 300 slab and 600 nm strip platforms and $n_{platform}=1.8-2.7$ for the 300 nm strip 
we get transmission larger than 80 \%, 81 \% and 94\% respectively ( see Fig. \ref{T_stacks_wl}).

\begin{figure}[!htb]
\centering\includegraphics[scale=1]{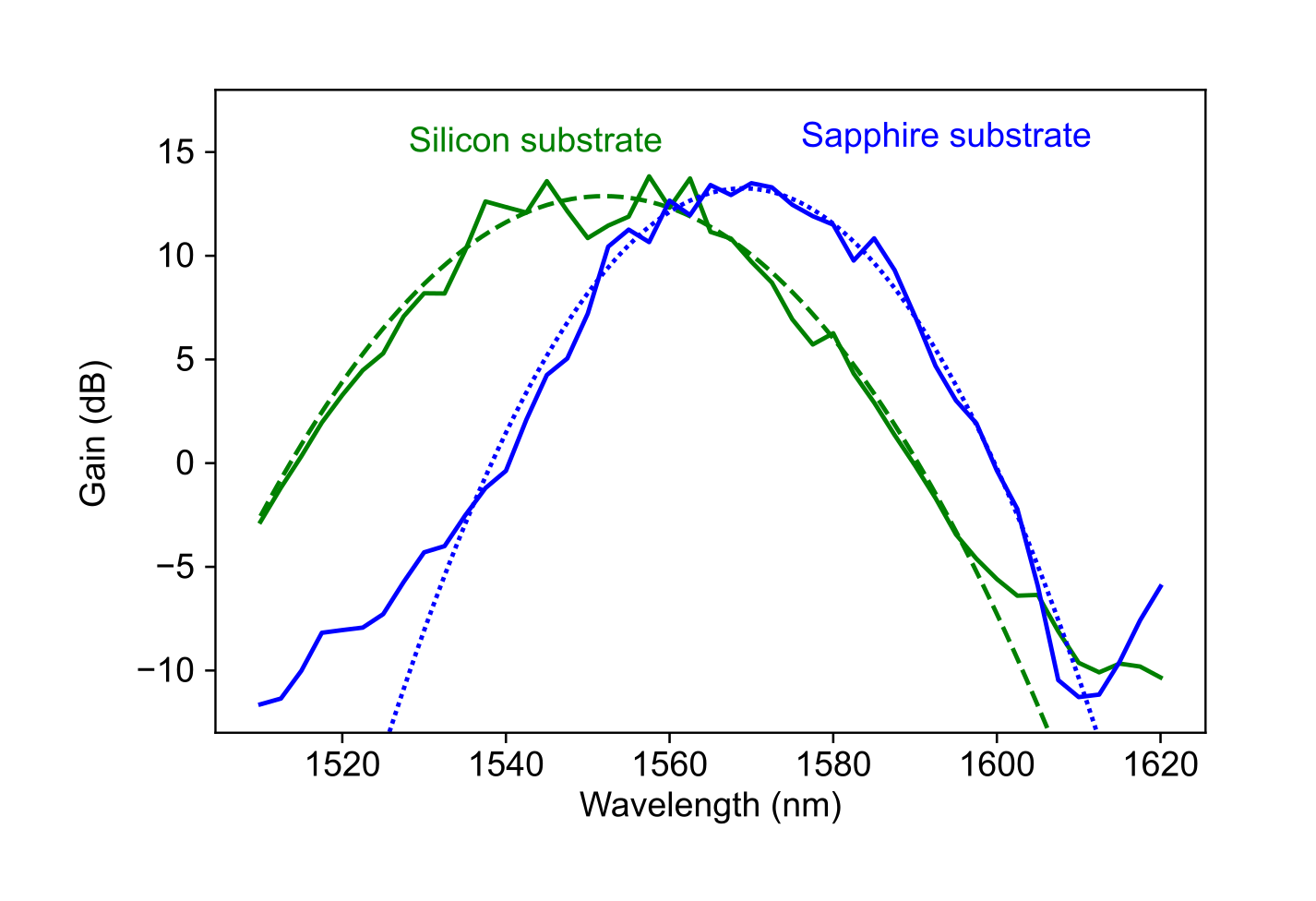}
\caption{\textbf{Dependence of the gain spectrum of amplifier coupon 3 on the substrate material:} Due to different thermal conductivity of the substrate material the internal temperature of the amplifier will be different for the same ambient temperature. Resulting in a wavelength shift of the gain peak. Here we compare amplifier 3 at 120 mA drive current printed on a LN on BOX on silicon ($k_T \approx 300 \frac{W}{m.K} $) to one printed on LN on BOX on sapphire ($k_T \approx 30 \frac{W}{m.K} $) and observe a shift of 16 nm in center wavelength. The dashed lines show the theoretical fit to the amplifier gain.
}
\label{amp_substrate}
\end{figure}
When considering changes in the BOX material or the substrate the simulations again indicate that the mode barely interacts with them so they don't affect the laser performance in first approximation.

In practice however the lasing wavelength will depend weakly on the substrate material. This effect is caused by different thermal conductivity ($k_T$) of the substrate leading to more or less self-heating of the amplifier. This then leads to a different temperature in the cavity, resulting in a shift of the gain spectrum of the amplifier and a change in the refractive index of the silicon grating which finally causes the lasing wavelength to shift.
 The amplifier gain shift is investigated by printing and measuring amplifiers on an LN on BOX on sapphire platform and comparing it to the LN on BOX on silicon platform considered in the main text. Fig. \ref{amp_substrate} shows the measured gains with a drive current of 120 mA on each platform with the theoretical fit of the small signal gain $g(\lambda)=Ge^{-A(\lambda-\lambda_0)^2} $ with $\lambda_0$ the center wavelength, $G$ the gain at the center wavelength and $A$ a phenomenological parameter. Even though the thermal conductivity of silicon ($k_T \approx 300 \frac{W}{m.K} $) is an order of magnitude larger than the conductance of sapphire ($k_T \approx 30 \frac{W}{m.K} $) the wavelength shift is only 16 nm . 
 
 To estimate the shift in the resonance wavelength of the silicon grating due to the change in the cavity temperature we look at measurements at different drive currents because here the effect is similar, a higher drive current induces more self-heating and shifts both the gain spectrum and the lasing wavelength. When sweeping the current for both the test amplifiers and the lasers we see a shift in lasing wavelength similar to the shift in the gain bandwidth. We thus estimate a shift of also $\sim 16$ nm in the lasing wavelength. To compensate for this shift a grating with a pitch 3 nm shorter would suffice. 
 
In practice these shifts can thus be anticipated as they result from a different substrate and can easily be ameliorated by having a large enough range of ready-made gratings and amplifier coupons. 
 We note that once the amplifiers and lasers are tested on a specific substrate, lasers can be printed on any platform with that substrate. The result of the test can be used to accurately predict the lasing wavelength regardless of the specific stack because of the platform agnostic design of the laser.

\section{Insensitivity of the lasing wavelength to the platform index}\label{sec:grating_wl_shift}
\begin{figure}[!hbt]
\centering\includegraphics[scale=1]{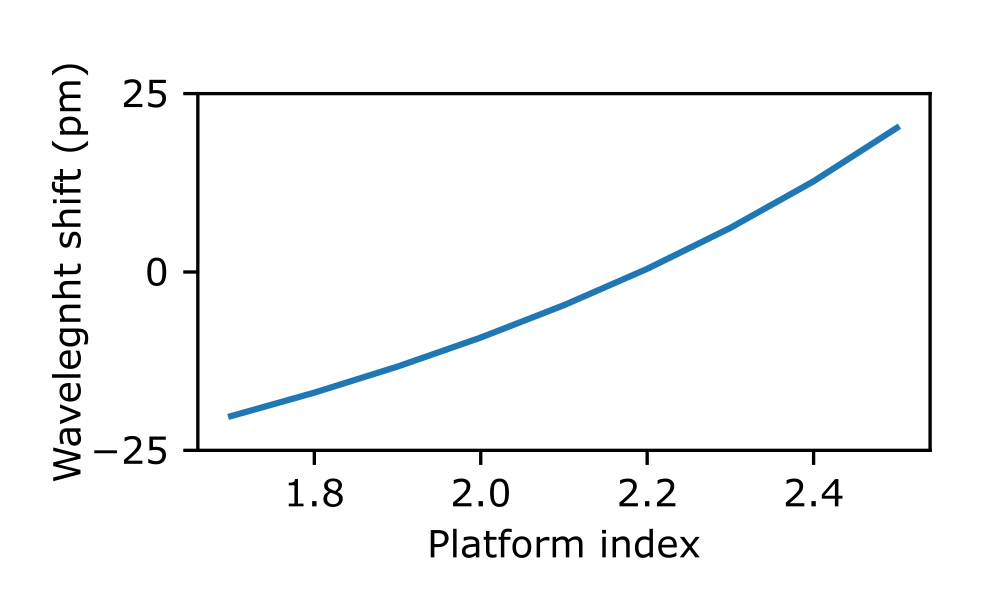}
\caption{\textbf{Cavity resonance wavelength dependence on the platform index} For a QWSDFB grating with a design wavelength of 1570 nm and grating pitch of 247 nm, the resonance wavelength only shifts 50 pm for platform indices ranging from 1.7 to 2.6. This is because the mode is almost completely confined in the silicon and III-V material and barely interacts with the platform material.
}
\label{grating_n_platform}
\end{figure}

As described in the main text, the mode inside the laser cavity is strongly confined in the silicon and III-V layers and barely interacts with the platform material. Fig. \ref{grating_n_platform} shows the resonance wavelength shift when varying the platform index between 1.7 and 2.6 for a QWSDFB grating with a design wavelength of 1570 nm and grating pitch of 247 nm. In our index range the resonance wavelength shifts only 50 pm, gratings with a different pitch have similar results. Aside from the caveat described in the previous section, the lasing wavelength selection can thus be done by choosing the corresponding silicon coupon pitch regardless of the target platform.

\section{Geometry of the coupons and the transfer printing site}\label{sec:geometry}
 In the main text we discuss the standardized process flow for platform agnostic laser integration. Here we describe in detail the geometry of the coupons and the transfer printing site on the target chip.
 
 The printing site on the target platform is defined by tapering the standard waveguide to a 3 $\mu m$ waveguide section, this section extends over 130 um, after which the waveguide tapers up to a width of 8 $\mu m$ over 30 $\mu m$, then it tapers out to a width of 18 $\mu m$ over 8 $\mu m$ which then abruptly transitions to a $>60$ \textmu m wide slab (for positive resist lithography no trenches are etched in this section leaving a full slab ). The wide slab will then continue for the length of the laser which is in this case $\sim$  1300 \textmu m and transition to the mirrored coupling section on the other side.  
 The 3 $\mu m$ wide section will be used to transfer the light to the silicon waveguide on top, the rest of the tapering is used to adiabatically transition to a slab and stay far enough from the silicon waveguide edges which are also tapering. The slab section is where the light will be completely confined in the silicon and III-V coupons on top. 
After defining the transfer printing site on the target platform and depositing the BCB adhesion and A\textsubscript{2}lO\textsubscript{3} etch stop layer, the silicon coupon can be printed. 
The silicon coupon is fabricated on a 400 nm Si layer and consists of a 1750 $\mu m \times$ 46 $\mu m$ rectangular coupon with a shallow etched (180 nm etch depth) $3\mu m$ wide waveguide across the coupon length and with a 450 $\mu m$ long shallow etched QWSDFB grating ($\Lambda=$ 247 nm) in the center. After printing, a fully etched taper is etched at both ends of the waveguide. Each silicon taper has a narrow tip of 120 nm which tapers out to 340 nm width over 120 $\mu m$ after which it tapers out to the 3 $ \mu m$ waveguide over a 30 $\mu m$ length. The tapers will be printed on the 3 $\mu m$ wide section of the transfer printing site on the target platform and transfer the light from the silicon waveguide to the platform.
The III-V coupon is fabricated with the stack described in supplementary section \ref{sec:amp_stack} , and has tapers with a nonlinear profile at each end to efficiently couple from light the cavity section to the 3 $ \mu m$ wide silicon waveguide. The III-V coupon can then be printed on top of the silicon coupon where the cavity is defined by the silicon QWSDFB but the optical mode is confined in both the silicon and the III-V gain section above. The details of the III-V coupon geometry is described in \cite{Haq:20amp}.
 \section{Robustness to transfer printing offsets}\label{sec:tp_offset}
\begin{figure}[!htb]
\centering\includegraphics[scale=1]{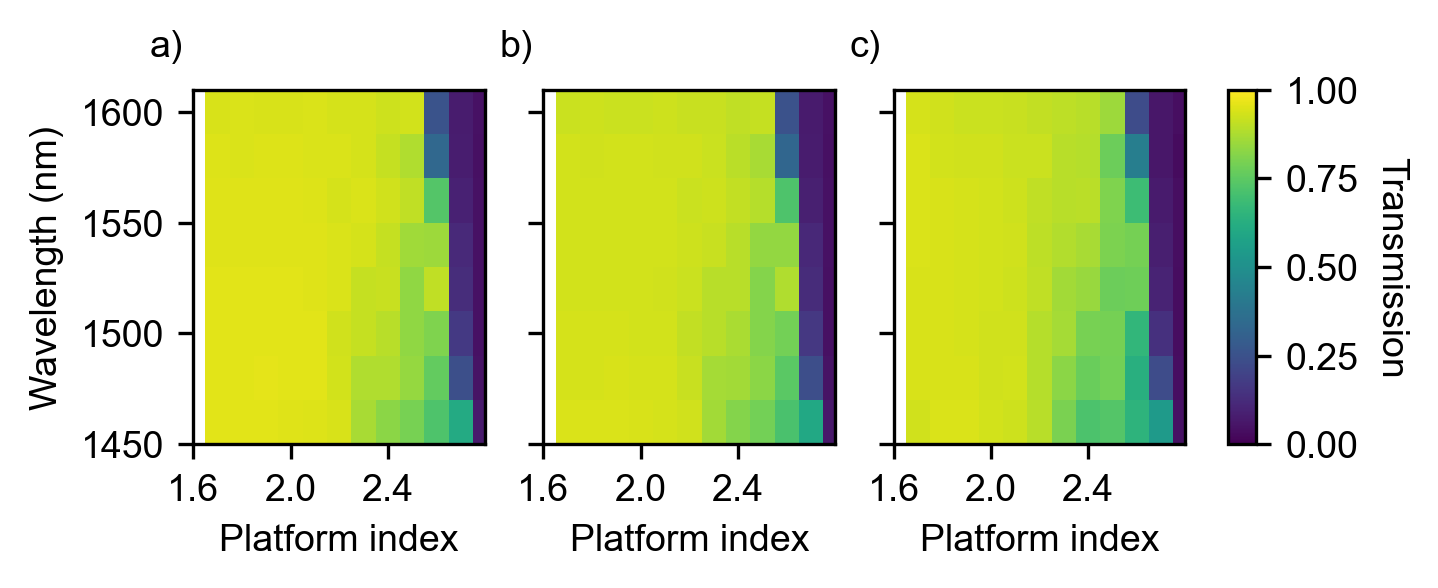}
\caption{\textbf{Power transmission from platform agnostic lasers to the platform for different transfer printing offsets} When considering transfer printing offsets of 500 nm for b) the silicon coupon and c) the amplifier coupon we see the double taper maintains a high transmission across different platform indices indices and wavelengths. The platform stack considered is the 600 nm rib waveguide with a 300 nm slab where a) shows the transmission with no offsets for comparison.}
\label{T_TPOffset}
\end{figure}

\begin{figure}[!htb]
\centering\includegraphics[scale=1]{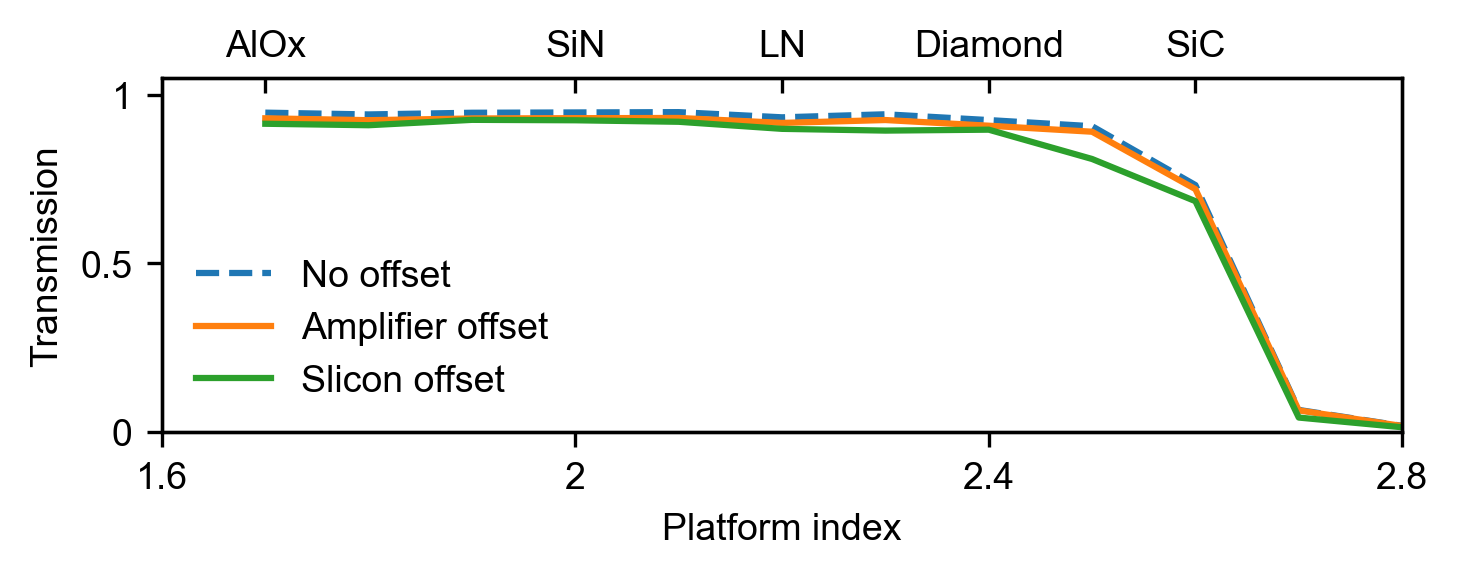}
\caption{\textbf{Power transmission from a platform agnostic laser at 1570 nm to the platform for different transfer printing offsets} When considering transfer printing offsets of 500 nm for the silicon coupon and the amplifier coupon we see the double taper maintains a high transmission across different platform indices indices and wavelengths. The platform stack considered is the 600 nm rib waveguide with a 300 nm slab where the transmission with no offsets is added for comparison.
}
\label{T1550_TPOffset}
\end{figure}

To check the robustness of our laser integration technique to lateral transfer printing offsets we consider an offset of 500 nm. This is three times the standard deviation of state of the art transfer printing tools and one time the standard deviation for our transfer printing tool $3\sigma^{SOA}=\sigma=$ 500 nm , this means that respectively 99.7\% and 68\% of the printed coupons will have less or equal offset for the different tools. We consider the offsets of the silicon and amplifier coupons separately and plot the simulated total transmission from the platform to the cavity in Fig. \ref{T_TPOffset} for different platform indices and wavelengths. For platform indices between 1.7 and 2.5 and wavelengths between 1460 nm and 1580 nm the transmission with an offset Si coupon is still larger than 74\% and for an offset amplifier coupon it is larger than 78\%, At any point across the opration domain this is a drop of less than 10 \% and 2\% respectively, compared to the perfectly aligned case in . The transmission mainly drops at the edge of our operation region however. Focusing on the transmission at 1570 nm we show in Fig. \ref{T1550_TPOffset} that the transmission only drops to $>80\%$ and $>89\%$ for the offset silicon and amplifier coupons respectively.

\section{Current tuning of the lasing wavelength on LN and SiN}\label{sec:tune lasers}
\begin{figure}[!htbp]
\centering\includegraphics[scale=1]{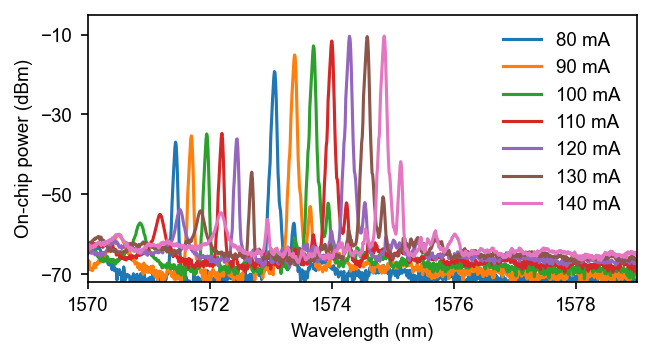}
\caption{\textbf{Experimental demonstration of tuning the wavelength of the platform agnostic laser on SiN using different drive currents:} Different drive currents will induce more or less self-heating of the laser resulting in a change in the refractive index of the QWSDFB grating ultimately shifting the wavelength of the laser. This can be used to tune the laser over $\sim$ 2 nm. 
}
\label{laser_tune}
\end{figure}

The lasers described in the main text are designed for a specific wavelength, but due to fabrication variations such as a different etch depth this wavelength can shift a bit. This is partly addressed by integrating the laser on a test platform and measuring the shift. The platform agnostic design then ensures that the shift will be near-identical on different platforms as systematic fabrication variations on the hundreds of coupons from the same chip will be the same. For some applications however e.g resonance based devices, the laser needs to emit at an exact wavelength. This is possible in our device by using current tuning to mitigate the shift from the design wavelength. Fig. \ref{laser_tune} shows that the wavelength can be tuned over $\sim$ 2 nm for currents going from 80 to 140 mA. The physical mechanism is that driving the amplifiers with more current induces more self-heating, changing the local temperature and the refractive index of the grating cavity, shifting the wavelength.
\section{Comparison of integrated lasers on LN and SiN}\label{sec:compare lasers}
\begin{figure}[!htbp]
 \centering \includegraphics[scale=1]{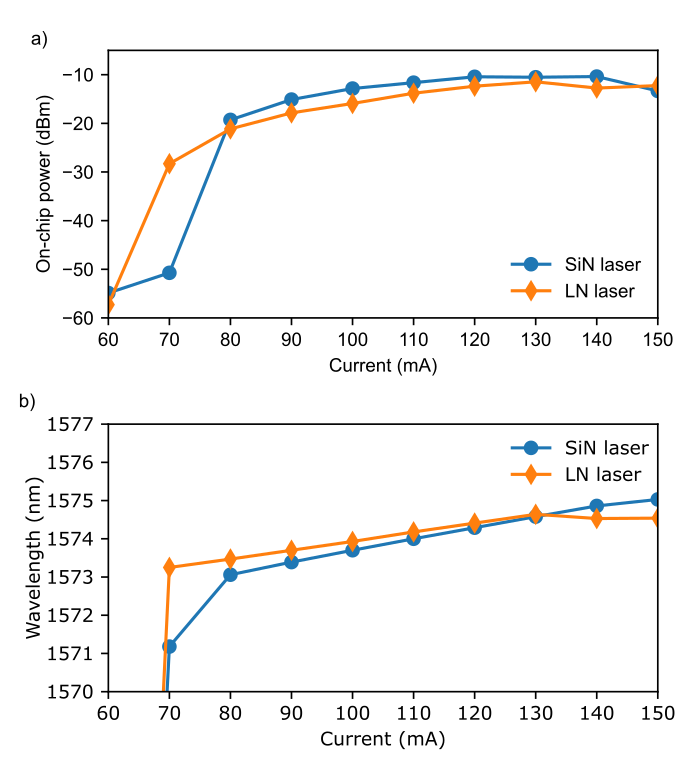}

 \caption{\textbf{Comparison of lasing characteristics of identical platform agnostic lasers printed on LN and SiN.} \textbf{a)} Lasing powers and \textbf{b)} lasing wavelengths for SiN and LN lasers for different drive currents. 
 }
 \label{laser_compare}
\end{figure}

In the main text platform agnostic lasers integrated on a SiN and LN platform are compared at 130 mA drive current. Here we compare them in both wavelength and power for their whole operating range. Fig. \ref{laser_compare} shows that the threshold currents are 70 mA and 80 mA for LN and SiN respectively. Moreover the lasing powers and wavelengths after threshold differ by mostly 3 dB and 0.5 nm respectively, at any drive current. This difference can be attributed to different temperature, chip coupling and other measurement conditions, fabrication variations should be minimal as the silicon coupons and III-V coupons are identical in design and come from the same chip. 


\newpage
\clearpage

\end{document}